\documentstyle[12pt]{article}
\begin{document}
\begin{titlepage}

\renewcommand\baselinestretch {1,5}
\large\normalsize

\begin{center}
Ukrainian Academy of Sciences\\
Institute for Condensed Matter Physics
\end{center}

\vspace{3cm}
\hspace{8cm}
\begin{tabular}{l}
Preprint\\
ICMP-94-9E
\end{tabular}

\vspace{3cm}

\begin{center}
{\large {O.Derzhko, T.Krokhmalskii\\}}
%\end{center}
\vspace{1cm}
%\begin{center}
{\Large {Statistical properties of random spin-$\frac{1}{2}$ $XY$ chains}}
\end{center}

\vspace{7cm}

\begin{center}
L'viv-1994\\
\end{center}

\end{titlepage}

\clearpage
\renewcommand\baselinestretch {1,5}
\large\normalsize

%\noindent
%\Ukrainian
%УДК 538.9\\
%\\
%О.Держко, Т.Крохмальський\\
%\\
%Статистичнў властивостў випадкових спўн-$\frac{1}{2}$ $XY$ ланцюжкўв\\
%\\
%Запропоновано числовий пўдхўд для вивчення рўвноважних статистичних
%властивостей спўн-$\frac{1}{2}$ $XY$ ланцюжкўв. Метод проўлюстровано
%дослўдженням впливу безладу на поперечну динамўчну сприйнятливўсть
%спўн-$\frac{1}{2}$ ланцюжка Ўзўнга у випадковому поперечному по\-лў.

\vspace{2cm}

\noindent
%\English
      O.Derzhko, T.Krokhmalskii \\
\\
Statistical properties of random spin-$\frac{1}{2}$ $XY$ chains\\
\\
A numerical approach to the study of equilibrium statistical properties of
spin-$\frac{1}{2}$ $XY$ chains is suggested. The approach is illustrated by
the examining of influence of disorder on transverse dynamical susceptibility
of spin-$\frac{1}{2}$ Ising chain in random transverse field.

\vspace{3cm}

\noindent
\copyright 1994 Institute for Condensed Matter Physics

\clearpage

\renewcommand\baselinestretch {1,5}
\large\normalsize
%\normalsize

One-dimensional spin-$\frac{1}{2}$ $XY$ models are among the few many-body
prob\-lems for
which a lot of exact results can be obtained. This is due to the fact that
the Hamiltonian of the system can be
rewritten via Jordan-Wigner transformation
in terms of non-interacting fermions \cite{lsm}.
In spite of this
there are rather limited results
for time-dependent
correlations between longitudinal spin components.
More difficult to deal with are the random versions of the model.
Apart from academic importance such results are used in the description of
a large number of quasi-one-dimensional ferroelectric and magnetic
compounds such as $Cs(H_{1-x}D_x)_2PO_4$, $PbH_{1-x}$ $D_xPO_4$, $PrCl_3$ etc.
In our previous
paper \cite{dk} we suggested a new numerical approach for evaluation of
thermodynamical properties and both equal-time and different-time spin
correlation functions
for spin-$\frac{1}{2}$ non-uniform ani\-so\-tro\-pic $XY$ chain in
transverse field with the Hamiltonian
\begin{eqnarray}
H=\sum_{j=1}^{N}\Omega_js_j^z+\sum_{j=1}^{N-1}(J^{xx}_js^x_js^x_{j+1}+
J^{yy}_js^y_js^y_{j+1}).
\end{eqnarray}
This method allows to study the random versions of the model (1) assuming
that ${...,\Omega_j,...;...,J^{xx}_j,...;...,J^{yy}_j,...}$ vary randomly from
site to site according to the given probability distribution.
The purpose of the present paper is to describe in more detail
the proposed approach and to apply it for
examining the influence of disorder on transverse dynamical susceptibility of
spin-$\frac{1}{2}$ Ising chain in random transverse field.
Such results, up to the best authors' knowledge,
are obtained for the first time.

In order to obtain the equilibrium statistical properties of the system in
question (1) one should first perform Jordan-Wigner transformation from the
raising and
lowering operators $s^\pm _j \equiv s^x_j\pm is^y_j$ to Fermi operators
$c_1=s^-_1, c_j=
\prod_{n=1}^{j-1}(-2s_n^z)s^-_j, j=2,...,N,$
$c^+_1=s^+_1, c_j^+=\prod_{n=1}^{j-1}(-2s_n^z)
s^+_j, j=2,...,N$
and then to diagonalize the obtained quadratic form by a canonical
transformation $\eta_k=\sum_{j=1}^N(g_{kj}c_j+h_{kj}c_j^+)$ with the outcome
$H=\sum_{k=1}^N\Lambda_k(\eta^+_k\eta_k-\frac{1}{2})$. The elementary
excitation
spectrum $\Lambda_k$ and the coefficients $g_{kn} \equiv
(\Phi_{kn}+\Psi_{kn})/2, h_{kn} \equiv (\Phi_{kn}-\Psi_{kn})/2$ are determined
from
\begin{eqnarray}
\Lambda_k\Phi_{kn}=\sum_{j=1}^{N}\Psi_{kj}({\bf A}+{\bf B})_{jn},
\Lambda_k\Psi_{kn}=\sum_{j=1}^{N}\Phi_{kj}({\bf A}-{\bf B})_{jn};
\nonumber\\
A_{ij} \equiv \Omega_i \delta_{ij}+ \frac{J^{xx}_i+J^{yy}_i}{4}
\delta_{j,i+1} + \frac{J^{xx}_{i-1}+J^{yy}_{i-1}}{4} \delta_{j,i-1},
\nonumber\\
B_{ij} \equiv \frac{J^{xx}_i-J^{yy}_i}{4} \delta_{j,i+1} -
\frac{J^{xx}_{i-1}-J^{yy}_{i-1}}{4} \delta_{j,i-1}.
\end{eqnarray}
Eqs. (2) reduce to standard problems
\begin{eqnarray}
\Lambda_k^2\Phi_{kn}=
\sum_{j=1}^{N}\Phi_{kj}[({\bf A}-{\bf B})({\bf A}+{\bf B})]_{jn},
\nonumber\\
\Lambda_k^2\Psi_{kn}=
\sum_{j=1}^{N}\Psi_{kj}[({\bf A}+{\bf B})({\bf A}-{\bf B})]_{jn}
\end{eqnarray}
for $N\times N$ five diagonal banded
matrices $({\bf A}\mp{\bf B})({\bf A}\pm{\bf B})$.

Next it is necessary to express the quantities of interest
through  $\Lambda_k$, $\Phi_{kj}$, $\Psi_{kj}$. One immediatelly notes that the
evaluation of partition function $Z(\beta , N)=
\prod_{k=1}^{N}2\cosh\frac{\beta\Lambda
_k}{2}$ involves all $\Lambda_k^2$.
Thus, the knowledge of eigenvalues of
matrices $({\bf A}\mp{\bf B})({\bf A}\pm{\bf B})$
permits one to obtain the thermodynamical properties of the model in
question (1).
Somewhat more complicated is the investigation of spin correlation functions.
Reformulating spin operators in terms of operators $\varphi_j^{\pm} \equiv
c^+_j \pm c_j$
(they are the known linear combinations of operators $\eta_k,
\eta_k^+$, namely,
$\varphi_j^+ = \sum_{m=1}^N \Phi_{mj} (\eta^+_m + \eta_m)$,
$\varphi_j^- = \sum_{m=1}^N \Psi_{mj} (\eta^+_m - \eta_m)$),
that is, $s_j^x= \prod_{n=1}^{j-1}(\varphi_n^+ \varphi^-_n)
\varphi^+_j /2$,
$s_j^y= \prod_{n=1}^{j-1}(\varphi_n^+ \varphi^-_n) \varphi^-_j /(2i)$,
$s_j^z=- \varphi_j^+ \varphi^-_j /2$,
noting that the calculation of spin correlation function reduces to
exploiting Wick-Bloch-de Dominicis theorem
and calculating the elementary contractions
\begin{eqnarray}
<\varphi^+_j(t) \varphi^+_m>=\sum_{p=1}^N \Phi_{pj} \Phi_{pm}
\frac{\cosh (i\Lambda_p t - \frac{\beta \Lambda_p}{2})}
{\cosh \frac{\beta \Lambda_p}{2}},
\nonumber\\
<\varphi^+_j(t) \varphi^-_m>=\sum_{p=1}^N \Phi_{pj} \Psi_{pm}
\frac{\sinh (-i\Lambda_p t + \frac{\beta \Lambda_p}{2})}
{\cosh \frac{\beta \Lambda_p}{2}},
\nonumber\\
<\varphi^-_j(t) \varphi^+_m>=-\sum_{p=1}^N \Psi_{pj} \Phi_{pm}
\frac{\sinh (-i\Lambda_p t + \frac{\beta \Lambda_p}{2})}
{\cosh \frac{\beta \Lambda_p}{2}},
\nonumber\\
<\varphi^-_j(t) \varphi^-_m>=-\sum_{p=1}^N \Psi_{pj} \Psi_{pm}
\frac{\cosh (i\Lambda_p t - \frac{\beta \Lambda_p}{2})}
{\cosh \frac{\beta \Lambda_p}{2}},
\end{eqnarray}
one finds that
$\Phi_{kj}$, $\Psi_{km}$ for all $\Lambda_k$ are involved in final
expressions for both equal-time and different-time spin correlation functions.
Therefore, the study of equilibrium statistical properties
may be
performed numerically in a following way:
first, to solve the eigenvalues and eigenvectors problem for matrix
$({\bf A}+{\bf B})({\bf A}-{\bf B})$ (3) obtaining in result
$\Lambda_k^2$ and $\Psi_{kj}$, and second, having
$\Psi_{kj}$ and
$\Lambda_k=\sqrt{\Lambda_k^2}$, to find $\Phi_{kj}$ from Eq.(2);
thermodynamics and spin correlation functions are expressed via
the sought quantities.
It should be stressed
that here appears only the eigenvalues and eigenvectors problem for
$N\times N$
five diagonal banded
matrix that is the remarkable
peculiarity of the system in question (1).
This is a key difference of our approach in comparison with other
finite-chain calculations (reported e.g. in \cite{dgs}) where the numerical
diagonalization of the Hamiltonian (1) in the Hilbert space of
dimension $2^N$ is performed (see also \cite{rc,gkpmtb}).

We end up with presenting the results of numerical calculation of
transverse dynamical susceptibility
%\begin{eqnarray}
$
\chi_{zz}(\ae, \omega) \equiv
\sum_{n} e^{i \ae n}
\int_0^{\infty} dt e^{i( \omega +i \epsilon ) t}
\frac{1}{i} <[s^z_j(t),$ $s^z_{j+n}]>
$
%\end{eqnarray}
for spin-$\frac{1}{2}$ Ising chain
$(J^{xx}_j=J, J_j^{yy}=0)$
in random transverse field
$\Omega_j$ with
probability distribution density
%\begin{eqnarray}
$
p(...,\Omega_j,...)=\prod_{j=1}^N[x\delta (\Omega_j)+
(1-x)\delta (\Omega_j-\Omega)]
$.
%\end{eqnarray}
Acting as described above one finds that
$
4<s^z_j(t)s^z_{j+n}>=< \varphi_j^+(t) \varphi^-_j(t)
\varphi_{j+n}^+ \varphi^-_{j+n}>=
< \varphi_j^+\varphi^-_j><\varphi_{j+n}^+ \varphi^-_{j+n}>
-< \varphi_j^+(t) \varphi^+_{j+n}><\varphi_{j}^-$ $(t) \varphi^-_{j+n}>
+< \varphi_j^+(t) \varphi^-_{j+n}><\varphi_{j}^-(t) \varphi^+_{j+n}>
$,
$<s^z_{j+n}s^z_j(t)>=<s^z_{j+n}(-t)s^z_j>$.
Computing then numerically the elementary contractions (4) involved in the
spin correlation functions
$<s^z_{100}(t)s^z_{100+n}>$,
$<s^z_{100+n}(-t)s^z_{100}>$
for the
random chain of $200$ spins
with $J=-2$, $\Omega =1$ at $\beta =20$,
performing numerically the integration over $t$ and
summation over $n$,
and averaging over the realization (typically few hundred) we
have obtained the results some of which are
presented in Fig.1. They agree with generally
known results for perfect (non-random) cases of
Ising model when $x=1$ and Ising model in transverse
field when $x=0$ derived for the first time
with periodic boundary conditions imposed
in \cite{n}.
As the concentration of the sites with transverse field $\Omega $ increases
$\chi_{zz}(\ae, \omega)$
rebuilds from the Ising-like type behaviour to the behaviour inherent to
Ising model in transverse field.
The divergence of Re$\chi_{zz}(\ae,\omega)$
at $\omega =2$ becomes finite, decreases, loses its symmetry and develops
into a smooth curve in the limiting case $x=0$.
The $\delta$-singularity at $\omega =2$ of Im$\chi_{zz}(\ae,\omega)$
transforms into few sharp cusps, loses its symmetry moving slightly to the
right and also develops into a smooth curve at $x=0$.
The study of the dependence of presented results on
number of spins in the chain, on upper limit in integration over $t$,
on value of $\epsilon$, on number of terms in sum over $n$
and on number of random realizations were performed carefully.
It should be noted that
the obtained susceptibilities exhibits a lot of structure (especially
at small $x$)
that requires further study.
%for essentially longer chains in order to make reliable extrapolations to
%$N=\infty $.

Since a complete solution for some pair time-dependent correlation
functions even in perfect spin-$\frac{1}{2}$ $XY$ chains has never
been found, the numerical calculations seem to be of much use in
deriving new results here. Some rigorous results on longitudinal
time-dependent correlation function were derived numerically for perfect
one-dimensional
spin-$\frac{1}{2}$ transverse Ising model at critical field and isotropic
$XY$ model in zero field at $\beta =\infty$ in \cite{ms},
however, within quite different approach.
The suggested method,
that gives exact results for finite spin-$\frac{1}{2}$ $XY$ chains,
appears to be of great use in
study of dynamical properties and random versions of systems considered.
The derived conclusions may be useful for interpretation of
observable data in dynamical
experiments on quasi-one-dimensional compounds that can be described in frames
of model (1).

The present paper was presented as a poster at International Conference on
Magnetism 1994 (22-26 August, 1994, Warsaw, Poland) (EP.227)
\cite{dk2}. The authors would like to express their gratitude to
Prof. L.L.Gon\c{c}al\-ves and other participants of ICM'94 for interest in
this paper and discussions.

\clearpage

\noindent
Fig. 1. Frequency-dependent real and imaginary parts of
transverse susceptibility
$\chi_{zz}(\ae ,\omega )$
for Ising chain in random transverse field
at $\ae =0$
for different values of concentration $x$;
$\epsilon =0.01$.

\vspace{6cm}
\renewcommand\baselinestretch {1}
\large\normalsize

\noindent
Fig. 1. Frequency-dependent real and imaginary parts of
transverse susceptibility
$\chi_{zz}(\ae ,\omega )$
for Ising chain in random transverse field
at $\ae =0$
for different values of concentration $x$;
$\epsilon =0.01$.

\clearpage

Address:\\
Institute for Condensed Matter Physics\\
1 Svientsitskii St., L'viv-11, 290011, Ukraine\\
Tel.: (0322)761054\\
Fax: (0322)761978\\
E-mail: icmp@sigma.icmp.lviv.ua

\end{document}